\documentstyle[twocolumn,aps,prl,epsf,epsfig]{revtex}

\begin{document}
\twocolumn[\hsize\textwidth\columnwidth\hsize
\csname@twocolumnfalse%
\endcsname
\draft
\title{Optically-Induced Polarons in Bose-Einstein Condensates:
Monitoring Composite Quasiparticle Decay}
\author{I.E.~Mazets$^1$\,$^,$\,$^2$, G.~Kurizki$^1$, N.~Katz$^3$,
and N.~Davidson$^3$}
\address{{\setlength{\baselineskip}{18pt}
$^1$\,Department of Chemical Physics, Weizmann Institute of Science,
Rehovot 76100, Israel,\\
$^2$\,Ioffe Physico-Technical Institute, St.Petersburg 194021, Russia, \\
$^3$\,Department of Physics of Complex Systems,
Weizmann Institute of Science, Rehovot 76100, Israel}}
\maketitle  
\begin{abstract}
Nonresonant light-scattering off atomic Bose-Einstein
condensates (BECs) is predicted to give rise to hitherto unexplored
composite quasiparticles: unstable polarons,
i.e., local ``impurities'' dressed by virtual phonons.
Optical monitoring of their spontaneous decay can display either Zeno
or anti-Zeno deviations from the Golden Rule, and thereby probe the
temporal correlations of elementary excitations in BECs.
\\     
\end{abstract}
\vskip1pc]

The evolution of a quantum state coupled to a 
continuum is coherent and effectively 
reversible over the so-called ``correlation time'' $t_{corr}$.  
This time marks the duration of the non-exponential initial 
stage of decay of the unstable state \cite{qz1,az1,az2}, 
or, equivalently, its 
response time to fast (impulsive) perturbations. 

Little is known about the short-time dynamics of 
impulsively-perturbed quantum many-body systems. 
Atomic Bose-Einstein condensates (BECs) are especially suitable for the 
exploration of such effects, since their 
elementary excitations, which are describable as quasiparticles 
\cite{kett,sbe,bcoe,rnb,bdae,cmod,rot}, may remain correlated 
for a long time. Here we consider the short-time dynamics 
of a composite quasiparticle, consisting of 
a moving ``impurity'' atom (an atom in a ground-state 
sublevel different from the rest of the 
BEC) ``dressed'' by a cloud of virtual phonons due to deformation 
of its vicinity. This quasiparticle may be viewed 
as the BEC analog of the solid-state polaron \cite{isp}. 
The time necessary for dressing an impurity atom by phonons is just the 
correlation time of the quasiparticle, 
which is shown to be exceedingly long ($\sim 1$~ms). 
The polaronic effect in a BEC is a genuine example of the response of  
a many-body quantum system  
to the presence of a probe particle. 
Direct measurement of $t_{corr}$ is a formidable task, since bare 
and dressed impurities are hardly distinguishable spectroscopically.

We therefore propose an indirect method of measuring $t_{corr}$:
optically-induced formation of polarons by Raman scattering
followed by real-time optical monitoring of the products after
their decay. The decay products are pair-correlated elementary
excitations that are {\em unambiguously} detectable.
Their time-resolved monitoring can reveal $t_{corr}$, which may be
manifest by either the Zeno or the anti-Zeno effects
\cite{qz1,az1,az2}, namely, slowdown or speedup of the decay rate
compared to its Golden Rule value. Fluctuations of the laser
fields, on time scales comparable to the very long $t_{corr}$ of
the BEC, in the ms range, can be used to interrogate these
features more easily. Similar results are obtained for the Bragg
process,  where, in contrast to the Raman process, the
atomic internal state is not changed.

\vspace*{16pt}

\begin{figure}
\begin{center}
\centerline{\epsfig{file=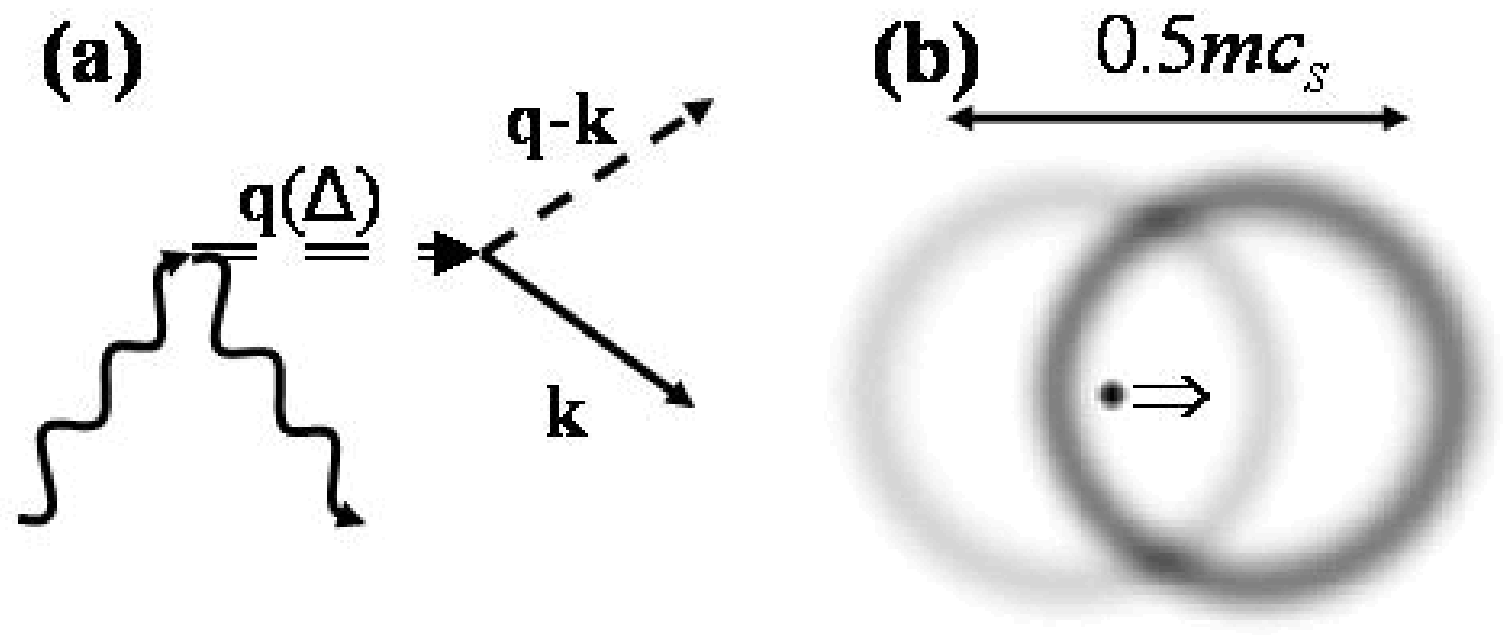,height=3.0cm}}
\end{center}
\begin{caption}
{(a) The Feynman diagram for the formation and decay of the
polaron includes the Raman beams (curved lines), the unstable
polaron (double dashed line) and the correlated products: the bare
impurity (dashed line) and BEC Bogoliubov excitation (solid line)
(b) The momentum distribution, (parallel and perpendicular to
$\hbar {\bf q}$, which is indicated by $\Rightarrow $) of the
impurity atoms (dark grey) and Bogoliubov excitations (light grey)
of the BEC (central dark spot) produced by Raman off-resonant
scattering with $\hbar \Delta =0.66\mu $, $\hbar q= 0.14 mc_s$.
The momentum shells are broadened due to finite pulse duration
($120 \hbar /\mu $). These shells should be observed in
time-of-flight images of the excited condensate.}
\end{caption}
\end{figure}

We shall primarily consider polaron formation by Raman scattering
off an atom in the BEC, as shown in Fig.~1\, (a). Two laser beams are
arranged such that momentum transfer of $\hbar {\bf q}$ is
accompanied by energy transfer of $\hbar \tilde{\Delta }$ to the
atom. The effective Rabi frequency associated with the induced
Raman two-photon transition is $\Omega $. Proper choice of the
laser-beam frequencies and polarizations ensures that the atom is
transferred to another sublevel of the ground state, thereby
leaving the condensate and forming an impurity atom 
\cite{brae1}. The energy of
such an impurity atom is given by its kinetic energy, $\hbar
^2q^2/(2m)$, plus the mean-field shift, $4\pi \hbar ^2an/m$, $a$
being the scattering length for a collision between the impurity
and condensate atoms, plus the energy-difference $E_D$ between the
two (Zeeman and/or hyperfine) sublevels involved. We can
incorporate the difference between the mean-field shift and  the
chemical potential $\mu =4\pi \hbar ^2a_0n/m$, $a_0$ being the BEC
scattering length, as well as $E_D$ into a new definition of the
Raman-transferred energy, $\hbar \Delta = \hbar \tilde{\Delta
}+4\pi \hbar ^2(a_0-a)n/m-E_D$. This enables us to identify the
energy of the impurity atom with its kinetic energy. We shall
assume large blue two-photon laser detuning, $\Delta   >\hbar
q^2/(2m)$, $\Omega \, ^<_\sim \, \Delta  $, so as to avoid the
need to deal with strongly-driven atoms \cite{scb}.

The relevant Hamiltonian for a uniform BEC is
\begin{equation}
\hat{H}=\hat{H}_{at}+\hat{H}_{int} ,
\label{H0}
\end{equation}
the first term describing the atomic system itself, and the second
one --- its interaction with the laser radiation. These terms are:
\begin{eqnarray}
\hat{H}_{at}&=&\sum _{\bf k}\hbar \omega _k \hat{b}^\dag _{\bf k}
\hat{b}_{\bf k}+ \sum _{\bf k}\frac{\hbar ^2k^2}{2m}
\hat{\beta }^\dag _{\bf k} \hat{\beta }_{\bf k}+ \nonumber \\ &&
\frac {4\pi \hbar ^2a\sqrt{n}}{m\sqrt{V}} \sum _{\bf k} \sqrt{S_k}
\left( \hat{\varrho }_{-{\bf k}}\hat{b}^\dag _{\bf k}
+ \hat{\varrho }_{\bf k}\hat{b} _{\bf k}  \right)  ,
\label{Hatom} \\
\hat{H}_{int}&=&\hbar \Omega \sqrt{N} \left( e^{-i\Delta  t}
\hat{\beta }^\dag _{\bf q}+e^{i\Delta  t}\hat{\beta }_{\bf q}\right) .
\label{Hint}
\end{eqnarray}
Here the creation and annihilation operators of the BEC elementary
Bogoliubov excitations are denoted by $\hat{b}^\dag _{\bf k},\,
\hat{b} _{\bf k}$, and those of impurity atoms by $\hat{\beta
}^\dag _{\bf k},\, \hat{\beta } _{\bf k}$, $\omega
_k=k\sqrt{[\hbar k/(2m)]^2+c_s^2}$ is the frequency of the BEC
elementary excitation with the momentum $\hbar k$, $c_s=\sqrt{\mu
/m}$ is the speed of sound in the BEC, $S_k=\hbar k^2/ (2m\omega
_k)$ is the BEC static structure factor, $\hat{\varrho }_{\bf k}=\sum
_{{\bf k}^\prime } \hat{\beta }^\dag _{{\bf k}+ {\bf k}^\prime
}\hat{\beta }_{{\bf k}^\prime }$ is the operator of the impurity
momentum shift, and $V$ is the quantization volume.

Now we define the dressed (polaronic) states as eigenstates of the
atomic-system Hamiltonian (\ref{Hatom}) and express the field-atom
interaction Hamiltonian (\ref{Hint}) in the basis of these states.
We omit the intermediate calculations, which are quite involved
but straightforward \cite{isp}, and present the result:
\begin{equation}
\hat{H}_{int} =\frac \hbar {\sqrt{V}}
\sum _{\bf k} \left[ {\cal M}_{{\bf q},\, {\bf k}}(t)
\hat{c}_{{\bf q}-{\bf k},\, {\bf k}}^\dag \hat{c}_0 +{\mathrm H.c.}
\right]  ,
\label{H123}
\end{equation}
where $\hat{c}_0$ is the operator of annihilation of a bosonic
atom in the BEC state, and $\hat{c}_{{\bf q}-{\bf k},\, {\bf
k}}^\dag $ is the operator of creation of a correlated pair
consisting of a dressed impurity atom with momentum $\hbar({\bf
q}-{\bf k})$ and a elementary Bogoliubov excitation with momentum
$\hbar{\bf k}$. The latter operator obeys the usual bosonic
commutation rules as long as the population of the state $|{\bf
q}-{\bf k},\, {\bf k}\rangle _d \equiv \hat{c}_{{\bf q}-{\bf k},\,
{\bf k}}^\dag |0\rangle $ is much less than 1. The energy of the
state $|{\bf q}-{\bf k},\, {\bf k}\rangle _d $ is $\epsilon
_{|{\bf q} -{\bf k}|,\, k}= \hbar \omega _k+\hbar ^2({\bf q}-{\bf
k})^2/(2m)$, provided that the energy correction (to second order
in $a$), which has to be calculated using the scattering-length
renormalization \cite{sbe}, is included in the definition of
$\hbar \Delta  $. The matrix element of the off-resonant
two-photon transition that couples the initial vacuum state
$|0\rangle $ to the state $|{\bf q}-{\bf k},\, {\bf k}\rangle _d $
is
\begin{equation}
{\cal M}_{{\bf q},\, {\bf k}}(t) =
\frac {4\pi \hbar a\sqrt{nS_{ k } }\Omega e^{-i\Delta t}}
{m[\, \omega _{k }+
\hbar k^{ 2}/(2m)-\hbar {\bf qk} /m \, ] }
\label{imp1a1}
\end{equation}
The relevant Feynman diagram is shown in Fig.~1\, (a). The operator
Heisenberg equations obtainable from Eq.~(\ref{H123}) are
\begin{eqnarray}
i\frac {\partial }{\partial t}
\hat{c}_0&=&\sum _{\bf k} {\cal M}_{{\bf q},\, {\bf k}}^*
\hat{c} _{{\bf q}-{\bf k},\, {\bf k}}, \label{eqa1} \\
i\frac {\partial }{\partial t}
\hat{c}_{{\bf q}-{\bf k},\, {\bf k}}&=&\epsilon _{|{\bf q}-
{\bf k}|,\, { k}} \hat{c} _{{\bf q}-{\bf k},\, {\bf k}}+
{\cal M}_{{\bf q},\, {\bf k}}\hat{c} _0 .   \label{eqa2}
\end{eqnarray}
Equations (\ref{eqa1}, \ref{eqa2}) represent the well-known model
of a discrete state coupled to a continuum. Assuming a
perturbative production fraction, the time-dependent rate of
creation of correlated pairs, consisting of an impurity atom and a
Bogoliubov excitation is
\begin{equation}
\Gamma (t)=\int \frac {d^3{\bf k}} {(2\pi )^3} \,
\left| {\cal{M}}_{{\bf q},\, {\bf k}} \right| ^2
\frac {2\sin (\nu t)}\nu  ,
\label{ww}
\end{equation}
where $\nu =\Delta -\omega _k -\hbar ({\bf q}-{\bf k})^2 /(2m)$.

In what follows we assume that $\hbar q\ll mc_s$, i.e., the {\em
virtual} impurity atoms move at a subsonic velocity. Whereas the
detuning of the beams $\Delta$ from the Raman resonance, ensures a
significantly larger momenta (and subsequent energy) for the decay
products.

The quantity $\hbar {\bf kq}/m$ can then be neglected compared to
$\omega _k$, and Eq.~(\ref{ww}) can be approximately expressed as
\begin{equation}
\Gamma (t)=\Gamma _*\int _0^\infty d\delta \, \frac {\sin [ \,
(\delta  - \Delta _q)t \, ]}{ \pi (\delta  - \Delta _q)} G(\delta
) .
\label{gammat}
\end{equation}
Here
\begin{equation}
\Gamma _{*}= 4\sqrt{\pi } \sqrt{\frac a{a_0}}
\frac {\hbar \Omega ^2}\mu \sqrt{na^3},
\label{novg}
\end{equation}
is the characteristic decay rate,
$G (\delta  )= (\hbar \delta /\mu )\, [ \, (\hbar \delta /\mu )+ 1 \, 
] ^{-5/2}$
is the dimensionless response function of the uniform BEC, and
$\Delta _q=\Delta  -\hbar q^2/(2m)\approx \Delta  $. The integral
in Eq.~(\ref{gammat}) can be performed analytically. The
proportionality of the rate prefactor $\Gamma _*$ in
Eq.~(\ref{novg}) to $\sqrt{na^3}$ clearly indicates that the
process under consideration relies on non-mean-field (impurity
scattering) effects. For $\mu t/\hbar \gg 1$ Eq.~(\ref{gammat})
gives the Golden Rule rate $\Gamma _{GR}=\Gamma _* G(\Delta _q)$.

In Fig. 2 we display $\Gamma (t)/\Gamma _{GR}$ for three different
detunings. For $\mu t/\hbar \, ^<_\sim \, 4$ the drastic
deviations from $\Gamma _{GR}$ imply that we are within the
correlation time $t_{corr}$ of the elementary excitations before
irreversibility sets in.

The Raman excitation fraction, namely, the number of correlated
pairs created per BEC atom, ${P}(t) = \int _0^t dt^\prime \,
\Gamma (t^\prime )$, is assumed to be small so as to avoid
condensate depletion: ${P}\ll 1$. 
Then Eq.~(\ref{gammat}) yields 
\begin{equation}
{P}(t) = \Gamma _*\int _0^\infty d\delta \, \frac {2\sin ^2 [ \,
(\delta  - \Delta _q)t /2\, ]}{ \pi (\delta  - \Delta _q)^2}
G(\delta  ) . \label{P1ppm}
\end{equation}

\begin{figure}
\begin{center}
\centerline{\psfig{file=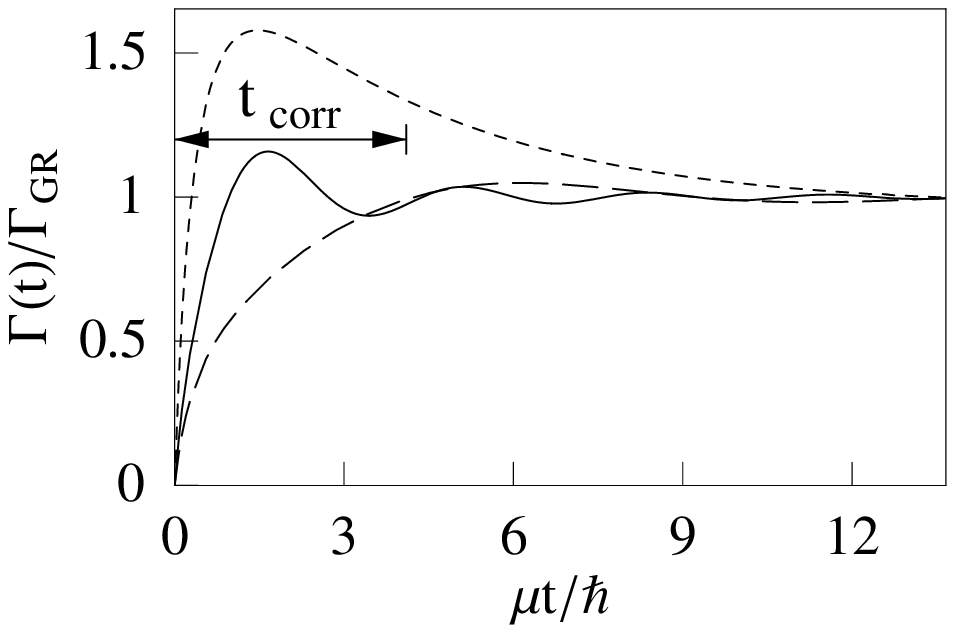,width=6.5cm}}
\end{center}
\vspace*{-0.6cm}
\begin{caption}
{The creation rate $\Gamma $ (normalized to $\Gamma _{GR}$) of the
laser-induced impurity atoms and Bogoliubov excitations as a
function of dimensionless time for $\hbar \Delta _q/\mu $ equal to
2.0 (solid line), 0.66 (long-dashed line), and 0.07 (short-dashed
line).  Typical correlation time is $t_{corr}$. }
\end{caption}
\end{figure}

\begin{figure}
\begin{center}
\centerline{\psfig{file=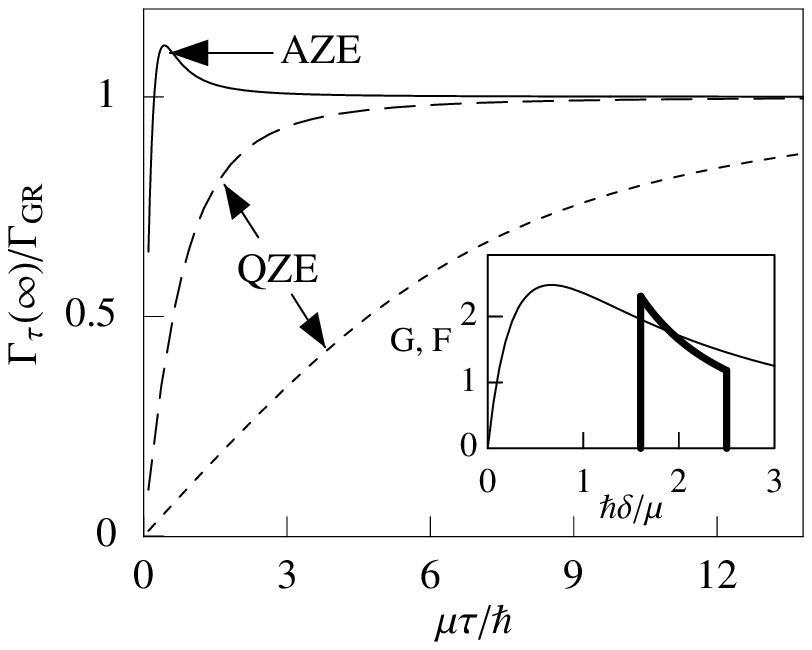,width=7.0cm}}
\end{center}
\vspace*{-0.6cm}
\begin{caption}
{The creation rate $\Gamma _\tau (\infty )$ of laser-induced
impurity atoms and Bogoliubov excitations
versus the dephasing time
$\tau $ for the same $\hbar \Delta _q/\mu $ as in Fig.~2.
AZE and QZE regions are indicated.
Inset: the medium response function $G (\delta )$
(thin line) and typical modulation spectrum $F(\delta )$ (thick line),
both in arbitrary units.}
\end{caption}
\end{figure}

The long correlation time of the BEC can lead to a significant
deviation of $P(t)$ from that expected from Golden Rule. In
principle, this deviation is observable, even at long times, as a
shift from the Golden Rule prediction, $P(t)=\Gamma _{GR}t$. 
However, 
this deviation is difficult to observe 
for realistic experimental parameters, as
 we show below.

We therefore suggest an alternative scheme, which 
allows us to observe large deviations
from the Golden Rule rate, even at the limit of large times.
Suppose that the difference of the two Raman laser frequencies is
randomly modulated \cite{az2}, so that the well-defined detuning
$\Delta _q$ is replaced by the spectral distribution $F(\delta )$
normalized to 1. Its spectral r.m.s. fluctuation (around the mean
frequency $\Delta _q$), $\tau ^{-1}$, may be regarded as the
inverse dephasing time. Under these conditions, even the long-time
limit of the excitation rate $\Gamma _\tau (\infty )$ obtained at
$t \gg \hbar /\mu $ may strongly differ from $\Gamma _{GR}$. As
follows from the universal formulas for the QZE  \cite{az2},
Eqs.~(\ref{gammat}, \ref{P1ppm}) take in this case the form
\begin{eqnarray}
\Gamma _\tau (t)&=&\Gamma _*\int _0^\infty d\delta
\int _0^\infty d\delta ^\prime \, \frac {\sin [ \,
(\delta  - \delta ^\prime )t \, ]}{ \pi (\delta  - \delta ^\prime )}
G(\delta )F(\delta ^\prime ) ,
\label{unfo} \\
P_\tau (t)& = &\Gamma _*\int _0^\infty d\delta
\int _0^\infty d\delta ^\prime \, \frac {2\sin ^2 [ \,
(\delta -\delta ^\prime )t /2\, ]}{ \pi (\delta -\delta ^\prime )^2}
G(\delta  ) F(\delta ^\prime )   .    \nonumber
\end{eqnarray}
Equation (\ref{unfo})   
implies that the modulation spectrum $F(\delta)$ controls
the asymptotic decay rate $\Gamma _\tau ( \infty ) $ and 
excitation fraction $P_\tau ( \infty )$, so that a suitable 
choice of the modulation spectrum may yield 
conspicuous deviations from the Golden Rule, in contrast to the case of  
Eq. (\ref{P1ppm}). 

Specifically, we choose 
the modulation spectrum $F(\delta )$ with a sharp 
low-frequency cut-off in order to suppress 
undesired transitions from
$|0\rangle $ to the state containing only a dressed impurity with
the momentum $\hbar {\bf q}$ and no real Bogoliubov excitations.
 These 
requirements are satisfied by  $F(\delta )= {\cal C}
\delta ^{-3/2}$ for $\delta _1<\delta <\delta _2$ and zero
otherwise (see inset of Fig.~3). The coefficient ${\cal C}$
ensures normalization to 1, and the cut-off frequencies $\delta
_1,~\delta _2$ are chosen so that the mean frequency is $\Delta
_q$ and the r.m.s. fluctuation is $\tau ^{-1}$.
 The resulting values 
for $\Gamma _\tau (\infty )$ are shown in
Fig.~3 as functions of $\tau $ for different values of $\Delta
 _q$. 

The convolution of $G(\delta )$ and $F(\delta )$ may give
rise to either the Quantum Zeno effect (QZE) or the opposite,
anti-Zeno effect (AZE), i.e. decay speedup at short times
\cite{az1}. As expected \cite{qz1,az1,az2}, the QZE always takes
place for extremely small times $\tau $. For large detunings, the
modulation spreads the accessible final states towards the range
where the response $G( \delta )$ {\em increases} as $\tau ^{-1}$
grows, and the irreversible transitions are accelerated (AZE takes
place) \cite{az2}. Alternatively, for small and moderate
detunings, the modulation spreads the final states over a range of
energies such that $G$ { \em decreases} as $\tau ^{-1}$ grows,
thereby bringing about the QZE. Comparison of Figs. 2 and 3 
indicates that the dephasing time scale  
$\tau $ associated with the QZE can be identified with the 
correlation time $t_{corr}$ of elementary excitations. 

We obtain qualitatively similar results for off-resonant Bragg
scattering, although there are some subtle caveats, due to the
indistinguishability of the excitations, as opposed to the Raman
process.  The general expression for the
rate $\Gamma _{Bragg}(t)$ of this process is quite cumbersome.
Here we present the $t\rightarrow \infty $ limit of $\Gamma
_{Bragg}(\infty )$ for the particular case $\Delta  \gg \mu /\hbar
\gg \hbar q^2/(2m)$:
\begin{equation}
\Gamma _{Bragg}(\infty )\approx 8\sqrt{\pi } \left(
\frac \mu {\hbar \Delta  }\right) ^{3/2} \frac {\hbar \Omega ^2}\mu
\sqrt{na_0^3}.
\label{B1}
\end{equation}
Remarkably, $\Gamma _{Bragg }$ does not depend on $q$, as long as
$q\ell \,^>_\sim \,1$, $\ell $ being the length of the trapped BEC 
sample. The Bragg process is characterized by a
correlation time, which is similar to that of the Raman process
and, hence, can be also revealed via the QZE and AZE.

\begin{figure}
\begin{center}
\centerline{\psfig{file=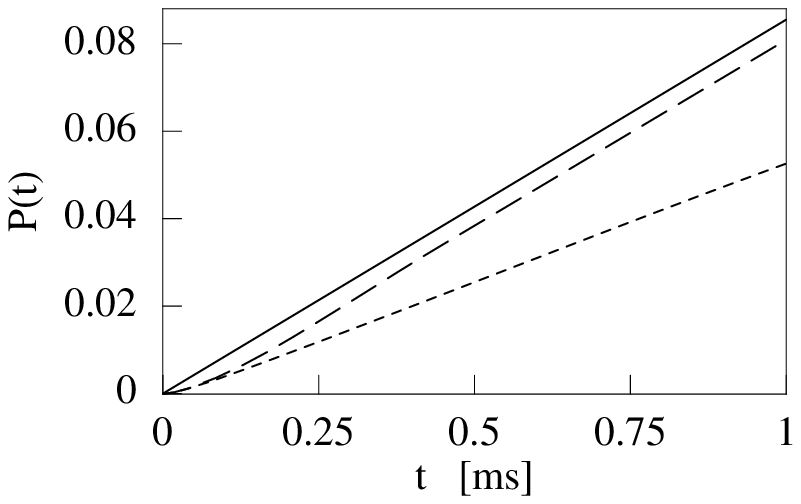,width=6.5cm}}
\end{center}
\vspace*{-0.6cm}
\begin{caption}
{The Raman excitation fraction (dimensionless),
as a function of time for a
$^{87}$Rb BEC with $n=4\cdot 10^{14}$~cm$^{-3}$  and
$\Omega =\Delta _q \approx 1.3\cdot 10^4$~s$^{-1}$. The solid line
is the Golden Rule result. The dashed line is the prediction for a
cw Raman experiment, and the dotted line is the prediction for a
frequency modulated Raman experiment, with the laser beam correlation
time $\tau =1.0\, \hbar /\mu $. The modulation leads to a clear
asymptotic
deviation from the Golden Rule result that is easily observed
experimentally, in contrast to the case of a cw Raman excitation. }
\end{caption}
\end{figure}

We used for our plots 
the response function $G(\delta )$  of a uniform BEC. 
It can be  generalized to the local density approach if the length
of the trapped BEC sample satisfies $q\ell
>1$ (usually $\ell \sim 10^{-3}$~cm). This inequality is
compatible with the subsonic limit for $\hbar q/m$ under typical
experimental conditions \cite{brae1,brae2}.

As an example, we consider a $^{87}$Rb BEC with $n=4\cdot
10^{14}$~cm$^{-3}$ \cite{bdae} and $\Omega =\Delta _q \approx
1.3\cdot 10^4$~s$^{-1}$, for the Raman process. We expect, after
 the 
excitation, to observe by the time-of-flight technique \cite{brae2} 
unambiguous momentum shells of Raman and
phonon excitations, as shown in Fig.~1\, (b). Although the probability
of direct excitation of dressed impurities with momentum $\hbar
{\bf q}$ may be non-negligible because of broadening of various
kinds (inhomogeneous, Fourier etc.), impurity atoms produced via
this channel are well-separated by energy and, therefore,
distinguishable by a time-of-flight measurement  from
those produced by the off-resonant Raman process, as indicated by
the double arrow in Fig.~1\, (b).

In Fig.~4 we show the total excitation fraction $P(t)$ (dashed
line), which should be contrasted with the Golden Rule production
of excitations (solid line), indicating the presence of the QZE.
Experimentally, measuring the population in the excitation shells
may allow us to resolve $t_{corr}$ by observing a shift in the
asymptotic increase of excitations. However, this is quite
challenging. Therefore, we propose to use the
frequency modulation scheme described above. This will lead to an
easily measured deviation of the excitation fraction from its 
Golden Rule counterpart for all times, as shown in the dotted
line, calculated for a laser beam correlation time $\tau=1.0\, \hbar /
\mu $. Specifically, after 1 ms this deviation is nearly $40\% $, and 
the corresponding 
excitation fraction $P_\tau \approx 0.05$ is easily detectable 
experimentally \cite{bdae,brae1}.

To summarize, we have presented a theory of off-resonant
stimulated Raman and Bragg scattering in BECs, followed by the
hitherto unexplored creation of correlated pairs of matter-wave
quanta - the decay of a composite polaron. These processes are
found to be highly suitable for experimental observation of the
quantum Zeno and anti-Zeno effects. These effects may be used as
unique probes of correlation times of the BEC response to fast
(impulsive) perturbations, and the onset of irreversibility in the
quasiparticle formation.

The support of the EC (the QUACS RTN and the ATESIT project),
ISF, the Israeli Ministry of Science,
and Minerva is acknowledged. I.E.M. also thanks the programs
Russian Leading Scientific Schools (grant 1115.2003.2) and 
Universities of Russia (grant UR.01.01.287).



\begin{thebibliography}{99}
\bibitem{qz1} B.~Misra and E.C.G.~Sudarshan, J. Math. Phys. {\bf 18},
756 (1977); L.~Fonda, G.C.~Girardi, and A.~Rimini,
Rep. Prog. Phys. {\bf 41}, 587 (1978).

\bibitem{az1} A.M.~Lane, Phys. Lett. A {\bf 99}, 359 (1983);
P.~Facchi, H.~Nakazato, and S.~Pascazio,
Phys. Rev. Lett. {\bf 86}, 2699 (2001).

\bibitem{az2} A.G.~Kofman and G.~Kurizki, Nature (London)
{\bf 405}, 546 (2000); Phys. Rev. Lett.
{\bf 87}, 270405 (2001); {\em ibid.} {\bf 93}, 130406 (2004).


\bibitem{kett} W. Ketterle and S. Inouye, C.R. Acad. Sci., Ser. IV
Phys. Astrophys. {\bf 2}, 339 (2001).

\bibitem{sbe} S.T. Beliaev, Zh. Eksp. Teor. Fiz. {\bf 34}, 433 (1958).

\bibitem{bcoe} E.~Hodby, O.M.~Marag\`{o}, G.~Hechenblaikner, and
C.J.~Foot, Phys. Rev. Lett. {\bf 86}, 2196 (2001).


\bibitem{rnb} J.~Rogel-Salazar, G.H.C.~New, and K.~Burnett, J.~Opt.~B:
Quantum Semiclass. Opt. {\bf 5}, S90 (2003).

\bibitem{bdae} N.~Katz, J.~Steinhauer, R.~Ozeri, and N.~Davidson,
Phys. Rev. Lett. {\bf 89}, 220401 (2002);
R.~Ozeri, N. Katz, J. Steinhauer, E.~Rowen, and
N.~Davidson, Phys. Rev. Lett. {\bf 90}, 170401 (2003).

\bibitem{cmod} D.~O'Dell, S.~Giovanazzi, and G.~Kurizki, J.~Mod. Opt.
{\bf 50} 2655 (2003).

\bibitem{rot} D.H.J.~O'Dell, S.~Giovanazzi, and G.~Kurizki,
Phys. Rev. Lett. {\bf 90}, 110402 (2003).

\bibitem{isp} A.~Isihara, {\em Statistical Physics} (Academic Press,
NY, 1971), \S 14.6.

\bibitem{brae1} A.P.~Chikkatur et al., Phys. Rev. Lett. {\bf 85}, 483
(2000).


\bibitem{scb} M.O.~Scully and M.S.~Zubairy, {\em Quantum Optics}
(Cambridge, Cambridge University Press, 1997), \S 10.2.


\bibitem{brae2} J.~Steinhauer, R.~Ozeri,
N.~Katz, and N.~Davidson, Phys. Rev. Lett. {\bf 88}, 120407 (2002).

\end{thebibliography}
\end{document}